\documentclass[11pt]{article}
\usepackage{epsfig}

\newcommand{\BABARPubYear}    {01}

\newcommand{\BABARProcNumber} {73}
\newcommand{\SLACPubNumber} {9036}
\newcommand{\LANLNumber} {0000}

\input babarsym
\newcommand\prd[3]   {{{Phys.\ Rev.\ }{\bf D #1} (#2) #3}}

\newcommand\prl[3]   {{{Phys.\ Rev.\ Lett.\ }{\bf #1} (#2) #3}}
\newcommand\ptp[3]   {{{Prog.\ Theor.\ Phys.\ }{\bf #1} (#2) #3}}

\begin{document}
{\thispagestyle{empty}

\begin{flushright}
SLAC-PUB-\SLACPubNumber \\
\babar-PROC-\BABARPubYear/\BABARProcNumber \\
hep-ex/\LANLNumber \\
October, 2001 \\
\end{flushright}

\par\vskip 4cm

\begin{center}
  \Large \bf Penguin Mediated $B$ Decays at \Lbabar 
\end{center}
\bigskip

\begin{center}
  \large
  G. Mancinelli\\
 University of Cincinnati \\
 Physics Department  \\
 Cincinnati, OH 45221 \\
 E-mail: {\em giampi@slac.stanford.edu}\\
  (for the \lbabar\ Collaboration)
\end{center}
\bigskip \bigskip

\begin{center}
\large \bf Abstract
\end{center}
We report on preliminary results of searches for penguin mediated $B$
decays based on 20.7 fb$^{-1}$ of data collected at the $\Upsilon$(4S)
peak with the \babar\ detector
at PEP-II. The following branching fractions have been
measured:  
${\cal B}(B^+ \to \phi K^+) = (7.7^{+1.6}_{-1.4}\pm 0.8)\times 10^{-6}$,
${\cal B}(B^0 
\to \phi K^0) = (8.1^{+3.1}_{-2.5}\pm 0.8)\times 10^{-6}$, ${\cal B}(B^+
\to \phi 
K^{*+}) = (9.7^{+4.2}_{-3.4}\pm 1.7)\times 10^{-6}$, ${\cal B}(B^0 \to
\phi K^{*0}) = 
(8.7^{+2.5}_{-2.1}\pm 1.1)\times 10^{-6}$, ${\cal B}(B^+\to \omega \pi^+)
= (6.6^{+2.1}_{-1.8}\pm 0.7)\times 10^{-6}$, ${\cal B}(B\to \eta 
K^{*0}) = (19.8^{+6.5}_{-5.6}\pm1.7)\times 10^{-6}$, where the first error is
statistical and the second systematic. For several other modes we
report upper limits on their branching fractions; for example 
for the following flavor-changing neutral current decays,
${\cal B}(B\to K l^+ l^-) < 0.6 \times 10^{-6}$, ${\cal B}(B\to K^* l^+
l^-) < 2.5 
\times 10^{-6}$, at 90\% Confidence Level (C.L.).

\vfill
\begin{center}
Contributed to the Proceedings of the International Europhysics Conference 
On High-Energy Physics (HEP 2001),\\ 
7/12/2001---7/18/2001, Budapest, Hungary
\end{center}

\vspace{1.0cm}
\begin{center}
{\em Stanford Linear Accelerator Center, Stanford University, 
Stanford, CA 94309} \\ \vspace{0.1cm}\hrule\vspace{0.1cm}
Work supported in part by Department of Energy contract DE-AC03-76SF00515 and NFS grant PHY 9901568.
\end{center}

  \section{Introduction}
 
Flavor-changing neutral currents are forbidden at the tree level in
the Standard Model (SM), hence such processes are possible only through penguin
loops or suppressed tree amplitudes proportional to small couplings in hadronic
flavor mixing (CKM matrix~\cite{ckm}). These rare decays are interesting
because their rates and kinematics are in principle sensitive
to new heavy particles, predicted for example by supersymmetry models,
which can enter the loop. Furthermore, it is possible to use some of 
these modes to search
for direct $CP$-violation, measuring the CKM angle $\beta$ in
a penguin environment~\cite{sin2b} and compare it with results at the tree
level (from the charmonium modes). In the process, we can also test a
great number of models just by measuring the branching 
fractions of penguin mediated $B$ decays. 

The low rates of these decays and their large backgrounds 
make a high luminosity $B$-factory necessary
for their searches. The results presented here are
derived from 20.7 fb$^{-1}$ delivered in 1999 and 2000 by
PEP-II~\cite{pepii} at the 
$\Upsilon$(4S) peak and 2.6 fb$^{-1}$ off
peak (running at $\sim$40 MeV below the resonance energy), for a total of
$\sim$22.7 
million $B\bar B$ events. A detailed description of the \babar\ 
detector can be found elsewhere~\cite{babar}.  Charge conjugate states are
assumed throughout and branching fractions are averaged accordingly.

  \section{Analysis}

Much of the background in these rare decays can be reduced by exploiting
the good 
charged particle identification of \babar. This is crucial in analyses
involving $B$ meson decays with no charm in the final state. $dE/dx$
from the tracking devices provides good $K/\pi$
separation at $p<0.8$ $GeV/c$, while the response of the internally
reflecting ring 
imaging Cherenkov detector (DIRC) is excellent at higher
momenta. Here only the measurement of the Cherenkov angle is used
to discriminate $K$ and $\pi$ (with better than 3 $\sigma$ separation up
to 3.5 GeV/c), while for (lower momentum) composite
particle daughters, so-called kaon selectors are used. These selectors combine
information 
from all the relevant subsystems and have typical efficiencies of $80$-$90\%$
with very low ($\sim2$-$3\%$) $\pi\to K$ misidentification probability. The very
tight selection 
criteria applied to electrons  give an efficiency of $\sim$88\% with a
corresponding $\pi\to e$ misidentification probability of 
$\sim$0.15\%. The typical muon selection efficiency for momenta greater than 
$\sim$1 GeV/c is 60-70\%, with $\pi\to \mu$ misidentification probability of
$\sim$2\%. Performances of 
particle identification are tested on many 
control samples kinematically selected from data. 
Since the backgrounds are dominated by combinatorics from the continuum and
since the continuum topology is jetty compared to the isotropic
distribution of the signal, event shape variables are exploited to fight
this kind of background. The main ones are a Fisher discriminant (already
used at CLEO~\cite{cleo1}) and the cosine of the angle between the thrust axis
of the $B$ and the rest of the event. 

$B$ meson candidates are selected using either the beam energy-substituted
mass ($m_{ES} = \sqrt{(E_{exp})^2 - ({p_B})^2}$, where $p_B$ is the
momentum of the reconstructed $B$ and $E_{exp}$ the $B$ candidate expected
energy)
or the energy-constrained mass ($m_{EC}$, obtained via a kinematic fit of
the measured candidate four momentum in the $\Upsilon$(4S) frame with the
constraint $E^*_B = E^*_{beam}$), and 
the difference between the reconstructed $B$ candidate and the beam
energies ($\Delta E = E^*_B - E^*_{beam}$), 
where the stars indicate variables evaluated in the $\Upsilon$(4S) rest
frame. The  
latter depends on the mass hypothesis, hence it is a good discriminant for
different final states. All analyses use an unbinned maximum likelihood
fit to extract the signal yields, while event counting methods are used just
as cross checks. Most analyses select their cuts with their
signal regions blinded.

  \subsection{$B\to \phi K^{(*)}$}

\begin{figure}
\begin{center}
\epsfysize 3.5in
\epsfbox{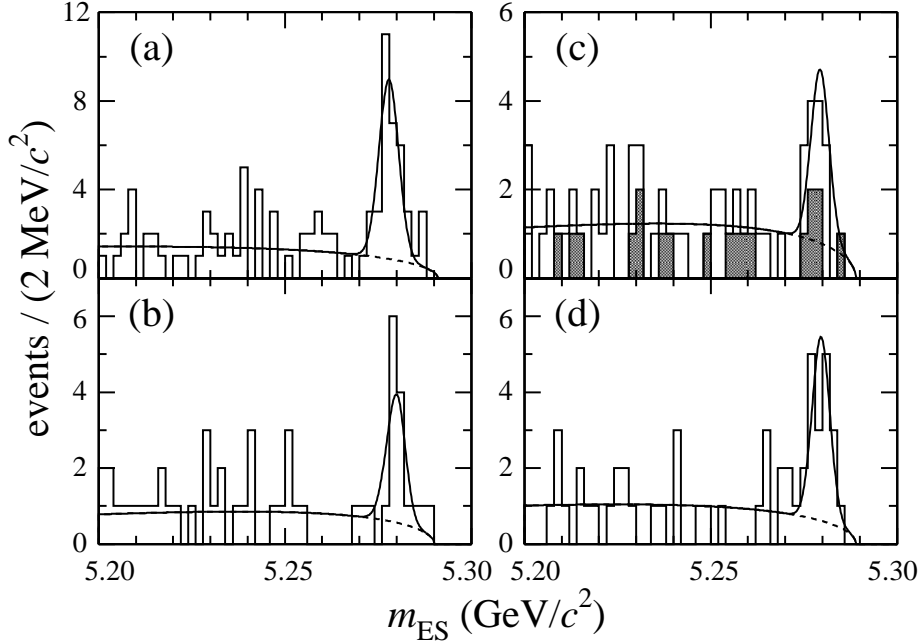}              
\caption{Projections onto the variable $m_{ES}$: (a) $B^+ \to \phi K^+$; (b) $B^0
\to \phi K^0$; (c) $B^+ \to \phi K^{*+}$; (d) $B^0 \to \phi K^{*0}$. In
(c) the histogram is the sum of the two $\phi K^{*+}$ channels, with the
shaded area corresponding to the $K^{*+} \to K^0_S \pi^+$ decay
chain.}
\label{phi1}
\end{center}
\end{figure}

\begin{table}
\begin{center}
\begin{tabular}{|l|l|} \hline 
Decay Mode& Branching Fraction \\
\hline 
\hline
$B^+\to \phi K^+$ & $(7.7^{+1.6}_{-1.4}\pm 0.8)\times 10^{-6}$ \\
$B^0\to \phi K^0$ & $(8.1^{+3.1}_{-2.5}\pm 0.8)\times 10^{-6}$ \\
$B^+\to \phi K^{*+}$ & $(9.7^{+4.2}_{-3.4}\pm 1.7)\times 10^{-6}$ \\
$B^0\to \phi K^{*0}$ & $(8.7^{+2.5}_{-2.1}\pm 1.1)\times 10^{-6}$ \\
$B^+\to \phi \pi^+$ & $<1.4\times 10^{-6}@90\%$ C.L. \\
\hline 
\hline
\end{tabular}
\caption{Results of branching fraction measurements for the $B\to \phi
K^{(*)}$ modes.}
\label{table1}
\end{center}
\end{table}

In the $B\to \phi K^{(*)}$ modes we reconstruct the final states $\phi
K^+$, $\phi K^0$, 
$\phi K^{*+}$ and $\phi K^{*0}$  where the following
intermediate states are recovered: $K^{*+} \to K^0 \pi^+$,  $K^{*+} \to
K^+ \pi^0$, $K^{*0} \to K^+ \pi^-$, $\phi \to K^+K^-$, $K^0 \to K^0_S
\to \pi^+\pi^-$ and $\pi^0 \to \gamma \gamma$. 
These decays are particularly interesting
because they are dominated by $b\to s(d)\bar s s$ penguins, with gluonic
and electroweak contributions, while other SM contributions
are strongly suppressed~\cite{sm}. In Figure~\ref{phi1} the
$m_{ES}$ distributions for the selected candidates for all the previous
modes are shown.  The solid (dashed) line shows the PDF projection of the
full fit (background only). The fits are mostly
for illustration purposes, since the branching fraction results are derived,
not from these, but from the maximum likelihood fits. A clear signal is
visible in all channels and we report a first observation of the decays
$B^+\to \phi K^{*+}$ and $B^0\to \phi K^0$. The final results for these modes
are reported in Table~\ref{table1}. They are consistent with
previously reported measurements~\cite{cleo2} and, within errors,
are consistent with isospin 
invariance under the assumption of penguin diagram dominance.
         
\begin{figure}
\begin{center}
\epsfysize 3in
\epsfbox{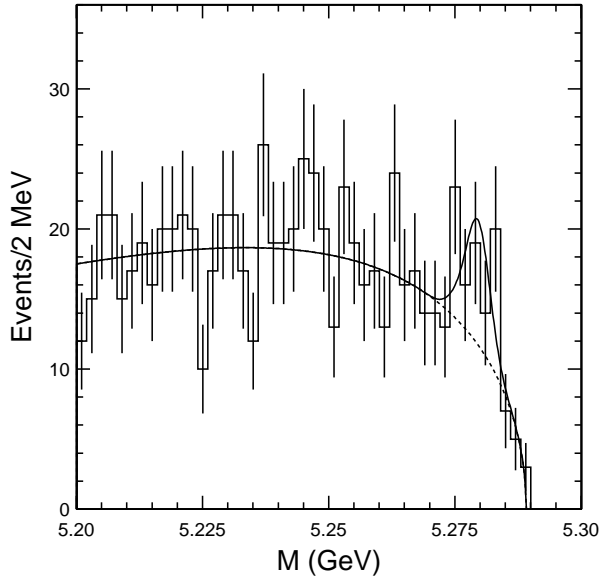}              
\caption{Projections of $m_{EC}$  for 
the $B\to \omega \pi^+$ mode. The projection is 
made by selecting events with signal likelihood (computed without
$m_{EC}$) exceeding a threshold that optimizes the expected
sensitivity.}
\label{omega1}          
\end{center}
\end{figure}

  \subsection{$B \to \omega X$}

$B$ mesons are reconstructed also in their decays into $\omega \pi^+$,
$\omega \pi^0$, $\omega K^+$, $\omega K^0_S$, with the $\omega$ decaying
into three pions. A signal is found solely for the mode $B\to \omega \pi^+$
(Figure~\ref{omega1}) and
the branching fraction measurement reported in Table~\ref{table2} is consistent
with those previously reported~\cite{cleo3}; in particular we find ${\cal
B}(B^+\to \omega \pi^+)> {\cal B}(B^+\to \omega K^+)$, as expected. Improved upper limits are
found for all the other decay channels. 

\begin{table}
\begin{center}
\begin{tabular}{|l|l|} \hline 
Decay Mode& Branching Fraction \\
\hline 
\hline
$B^+\to \omega K^+$ & $< 4\times 10^{-6} $ $ @90\%$ C.L. \\
$B^+\to \omega \pi^+$ & $(6.6^{+2.1}_{-1.8}\pm 0.7)\times 10^{-6}$ \\
$B^0\to \omega K^0$ & $< 14\times 10^{-6} $ $ @90\%$ C.L. \\
$B^0\to \omega \pi^0$ & $< 4\times 10^{-6} $ $ @90\%$ C.L. \\
\hline 
\hline
\end{tabular}
\caption{Results of branching fraction measurements for the $B\to \omega X$
modes.} 
\label{table2}
\end{center}
\end{table}

  \subsection{$B \to \eta K^*$}

\begin{figure}
\begin{center}
\epsfysize 3in
\epsfbox{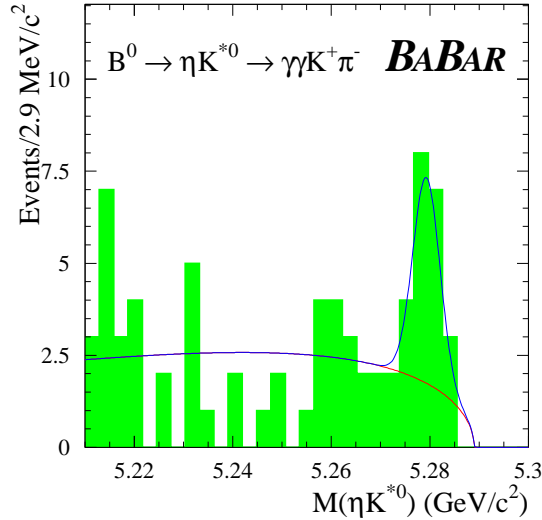}
\caption{$B$ candidate invariant mass for $B^0 \to \eta K^{*0}$. The histogram 
represents data and the smooth curve the fit function.}         
\label{eta2}        
\end{center}
\end{figure}

We have also analyzed the modes $B\to \eta K^{*+}$ and $B\to \eta
K^{*0}$, where the $\eta$ decays into two photons, the $K^{*0}$ 
into $K^+\pi^-$, the $K^{*+}$ into $K^0_S \pi^+$ and the $K^0_S$ into
$\pi^+\pi^-$. The $m_{EC}$  
projection plot is shown in Figure~\ref{eta2} for the   $\eta
K^{*0}$ channel, where a clear signal
is present and for which we measure a branching fraction of
$(19.8^{+6.5}_{-5.6}\pm1.7)\times 10^{-6}$. We
also find an upper limit of $33.9\times 10^{-6}$ for the branching fraction
of the $\eta K^{*+}$ 
mode at 90\% C.L. The results are 
consistent with those previously reported by the CLEO
Collaboration~\cite{cleo4} and, at least in 
the $\eta K^{*0}$ mode, confirm their rather high branching fraction
measurement.  

  \subsection{$B \to K^{(*)} l^+ l^-$}      

\begin{figure}
\begin{center}
\epsfysize 2.5in
\epsfbox{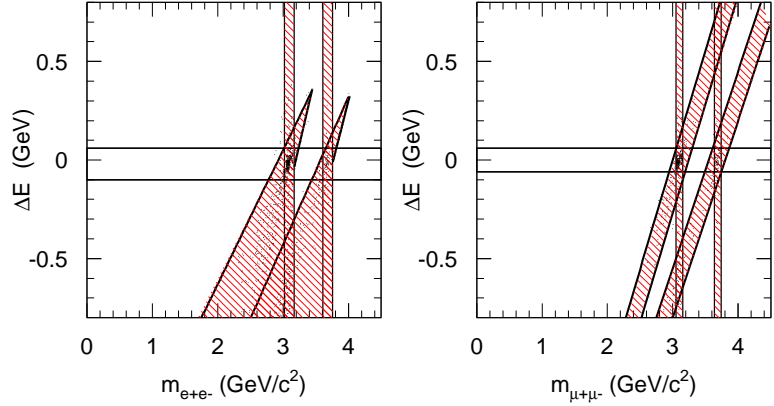}
\caption{The veto regions in the $\Delta E$ vs. $m_{l^+l^-}$ plane that are
populated by $B \to J/\psi K$ and 
$B \to \psi{\rm(2S)} K$ events.}
\label{kll5}        
\end{center}
\end{figure}
          
The dominant contributions for the processes $B \to K^{(*)} l^+ l^-$
in the SM 
are the so-called electroweak radiative penguins. These decays have been
reconstructed in the modes where $l=e,\mu$, the $K^{*0}$ decays
into $K^+\pi^-$ 
and the $K^{*+}$ into $K^0_S (K^0_S \to \pi^+\pi^-) \pi^+$. These
processes have a  
very clean experimental signature, due to the presence of both a lepton
pair and a kaon in the final state. 

\begin{figure}
\begin{center}
\epsfysize 4in
\epsfbox{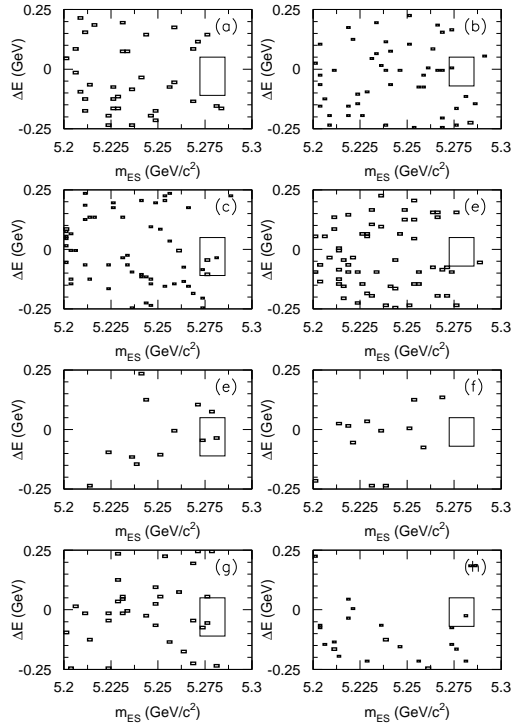}
\caption{Scatter plots of $\Delta E$ vs $m_{ES}$ after all analysis cuts
in  (a) 
$B^+ \to K^+
e^+ e^-$, (b) $B^+
\to K+ \mu^+\mu^-$, (c) $B^0 \to K^{*0}
e^+ e^-$, (d) $B^0
\to K^{*0} \mu^+\mu^-$, (e) $B^0 \to K^0_S
e^+ e^-$, (f) $B^0 \to K^0_S \mu^+\mu^-$, (g)  $B^+ \to
K^{*+}(\to K^0_S\pi^+) e^+ e^-$, (h) $B^+ \to K^{*+}(\to K^0_S\pi^+)
\mu^+\mu^-$. The small rectangles indicate
the signal region, which is used only for optimizing event selection
criteria.}
\label{kll2}
\end{center}
\end{figure}

The main challenge lies in understanding and characterizing the
background. To avoid biases as much as possible, a blind analysis is
performed, where not only the signal region, but also the sidebands used
to determine the background and its normalization, are blinded. Several
control samples from data are used to verify the reconstruction
efficiencies as determined from Monte Carlo (MC) simulation. In  particular $B\to J/\psi K^{(*)}$ and $B\to \psi{\rm(2S)}
K^{(*)}$ events are used, since they have the same topology as the signal
for these modes. Since these decays also constitute a dangerous
background, quite complex cuts in the $\Delta E$ and the lepton pair mass
plane are defined to exclude these events, taking into
account bremsstrahlung effects; they are shown in
Figure~\ref{kll5}, where the shaded areas are vetoed. (The same veto is
applied to the $K^*$ modes.) 
For reference, the two horizontal lines bound the region in which most
signal events are found. The 
charmonium veto removes these backgrounds not only from the signal region,
but also from the 
sideband region, simplifying the description of the background in the
fits. 

\begin{table}
\begin{center}
\begin{tabular}{|l|l|} \hline 
Decay Mode& Branching\\
& Fraction\\
& upper limits\\
& $(@90\%$ C.L.)\\
\hline 
\hline
$B^+\to K^+\mu^+\mu^-$ & $< 1.3 \times 10^{-6} $\\
$B^+\to K^+ e^+ e^-$ & $< 0.9 \times 10^{-6} $\\
$B^0\to K^0\mu^+\mu^-$ & $< 4.5 \times 10^{-6} $\\
$B^0\to K^0 e^+ e^-$ & $< 4.7 \times 10^{-6} $\\
\hline
$B\to K l^+ l^-$ & $< 0.6 \times 10^{-6} $\\
\hline
$B^0\to K^{*0}\mu^+\mu^-$ & $< 3.6 \times 10^{-6} $\\
$B^0\to K^{*0} e^+ e^-$ & $< 5.0 \times 10^{-6} $\\
$B^+\to K^{*+}\mu^+\mu^-$ & $< 17.5 \times 10^{-6} $\\
$B^+\to K^{*+} e^+ e^-$ & $< 10.0 \times 10^{-6} $\\
\hline
$B\to K^* l^+ l^-$ & $< 2.5 \times 10^{-6} $\\
\hline 
\hline
\end{tabular}
\caption{Results of branching fraction measurements for the $B\to K^{(*)} l^+
l^-$ modes.}
\label{table3}
\end{center}
\end{table}

\begin{figure}
\begin{center}
\epsfysize 4in
\epsfbox{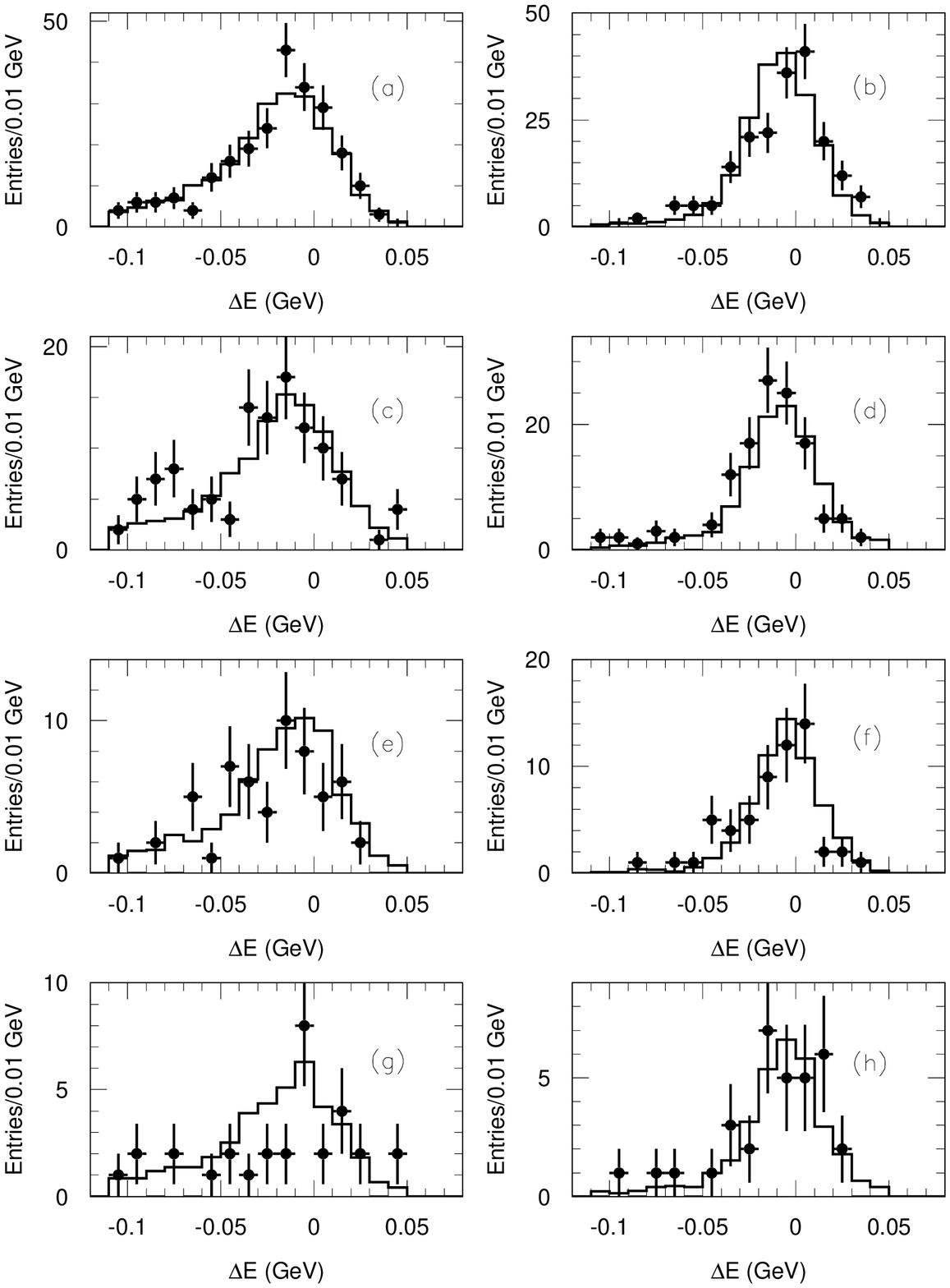}
\caption{ Comparison of $\Delta E$ shapes in several data control samples: (a) $B^+ \to
J/\psi(\to   
e^+ e^-)K^+$, (b) $B^+
\to J/\psi(\to \mu^+\mu^-)K+$, (c) $B^0 \to
J/\psi(\to
e^+ e^-)K^{*0}$, (d) $B^0
\to J/\psi(\to \mu^+\mu^-)K^{*0}$, (e) $B^0 \to
J/\psi(\to
e^+ e^-)K^0_S$, (f) $B^0
\to J/\psi(\to \mu^+\mu^-)K^0_S$, (g)  $B^+ \to
J/\psi(\to
e^+ e^-)K^{*+}(\to K^0_S\pi^+)$, (h) $B^+
\to J/\psi(\to \mu^+\mu^-)K^{*+}(\to K^0_S\pi^+)$.
The normalization is
absolute.}
\label{kll4}
\end{center}
\end{figure}

No excess of events in the signal regions is observed, as can be seen in
Figure~\ref{kll2}, where the $\Delta E$, $m_{ES}$ scatter
plots are shown for all decay modes. Very low level of
background is present. 
As a cross check, after
removing the $J/\psi$, $\psi{\rm(2S)}$ veto, we repeat the analysis on the
$J/\psi l l$ and $\psi{\rm(2S)} l l$ control samples. Not only is a clear
signal  
found, but the resulting $\Delta E$ shapes and normalizations are in very
good agreement with the simulation, as shown in Figure~\ref{kll4}, where
the points 
with error bars show the on-resonance data, and the solid histograms the
predictions of the 
charmonium MC simulation. All the analysis selection criteria have been
applied except for the 
charmonium veto, which is reversed. The
results are shown in 
Table~\ref{table3}. The combined limits for $B \to K^{(*)} l l$ are
now very close to the theoretical expectations for the SM~\cite{klltheo} and represent a significant improvement over previous
results~\cite{cdf}~\cite{cleo5}~\cite{belle}.

  \section{Summary}

With about 23 millions $B\bar B$ events \babar\ has already very
interesting results on penguin mediated decays. We have reported the first observation of 
the decays $B^+\to \phi K^{*+}$ and $B^0\to \phi K^0$ for which we found
the branching fractions $(9.7^{+4.2}_{-3.4}\pm 1.7)\times 10^{-6}$ and
$(8.1^{+3.1}_{-2.5}\pm 0.8)\times 10^{-6}$ respectively, and the upper
limit set 
on the ${\cal B}(B 
\to K l l$), $ < 0.6 \times 10^{-6}$, is very close to the theoretical
expectations for the 
SM. Other branching fractions have been
measured for penguin mediated $B$ decays: ${\cal B}(B^+ \to \phi K^+) =
(7.7^{+1.6}_{-1.4}\pm 0.8)\times 
10^{-6}$,  ${\cal B}(B^0 \to \phi K^{*0}) = 
(8.7^{+2.5}_{-2.1}\pm 1.1)\times 10^{-6}$, ${\cal B}(B^+\to \omega \pi^+) =
(6.6^{+2.1}_{-1.8}\pm 0.7)\times 10^{-6}$,  ${\cal B}(B\to \eta
K^{*0}) = (19.8^{+6.5}_{-5.6}\pm1.7)\times 10^{-6}$.

\section{Acknowledgments}

We are grateful to our PEP-II colleagues for their painstaking
dedication and tireless effort, which has made possible the achievement of
excellent luminosity and, consequently, the results presented here. The
collaborating institutions wish to thank SLAC for its support and kind
hospitality.

\end{document}